\documentclass[lettersize,journal]{IEEEtran}
\usepackage{cite}
\usepackage{amsmath,amssymb,amsfonts}
\usepackage{graphicx}
\usepackage{textcomp}
\usepackage{xcolor}

\usepackage{booktabs}
\usepackage{bm}
\usepackage{url}
\usepackage[ruled,linesnumbered]{algorithm2e}

\usepackage{amsmath,amsfonts}
\usepackage{array}
\usepackage[caption=false,font=normalsize,labelfont=sf,textfont=sf]{subfig}
\usepackage{textcomp}
\usepackage{stfloats}
\usepackage{url}
\usepackage{verbatim}
\usepackage{graphicx}
\def\BibTeX{{\rm B\kern-.05em{\sc i\kern-.025em b}\kern-.08em
    T\kern-.1667em\lower.7ex\hbox{E}\kern-.125emX}}
\usepackage{balance}
\begin{document}

\title{PPFM: Image denoising in photon-counting CT using single-step posterior sampling Poisson flow generative models}
\author{Dennis Hein, Staffan Holmin, Timothy Szczykutowicz, Jonathan S Maltz, Mats Danielsson, Ge Wang and Mats Persson
\thanks{This work involved human subjects or animals in its research. Approval of all ethical and experimental procedures and protocols was granted by the Swedish Ethics Review Agency under Application Nos. 2020-04638 and 2021-01092 and by the UW Madison Institutional Review board under Application No. 2022-1043 and performed in line with relevant Swedish and US legislation.}
\thanks{This study was financially supported by MedTechLabs, GE HealthCare, the Swedish Research council (grant no. 2021-05103), and the Göran Gustafsson foundation (grant no. 2114). \textit{(Corresponding author: Dennis Hein.)}}
\thanks{D. Hein, M. Persson, and M. Danielsson are with the Department of Physics, KTH Royal Institute of Technology, Stockholm, Sweden and MedTechLabs, BioClinicum, Karolinska University Hospital, Stockholm, Sweden. S. Holmin is with the Department of Clinical Neuroscience, Karolinska Insitutet, Stockholm, Sweden and the Department of Neuroradiology, Karolinska University Hospital, Stockholm Sweden. T. Szczykutowicz is with the Department of Radiology, University of Wisconsin School of Medicine and Public Health, Madison, WI, USA. J. Maltz is with GE HealthCare. G. Wang is with the Department of Biomedical Engineering, School of Engineering, Biomedical Imaging Center, Center for Biotechnology and Interdisciplinary Studies, Rensselaer Polytechnic Institute, Troy, NY, USA.}
\thanks{}}
\maketitle

\begin{abstract}
Diffusion and Poisson flow models have shown impressive performance in a wide range of generative tasks, including low-dose CT image denoising. However, one limitation in general, and for clinical applications in particular, is slow sampling. Due to their iterative nature, the number of function evaluations (NFE) required is usually on the order of $10-10^3$, both for conditional and unconditional generation. In this paper, we present posterior sampling Poisson flow generative models (PPFM), a novel image denoising technique for low-dose and photon-counting CT that produces excellent image quality whilst keeping NFE=1. Updating the training and sampling processes of Poisson flow generative models (PFGM)++, we learn a conditional generator which defines a trajectory between the prior noise distribution and the posterior distribution of interest. We additionally hijack and regularize the sampling process to achieve NFE=1. Our results shed light on the benefits of the PFGM++ framework compared to diffusion models. In addition, PPFM is shown to perform favorably compared to current state-of-the-art diffusion-style models with NFE=1, consistency models, as well as popular deep learning and non-deep learning-based image denoising techniques, on clinical low-dose CT images and clinical images from a prototype photon-counting CT system. 
\end{abstract}


\section{Introduction}
\IEEEPARstart{C}{omputed} tomography (CT) is a widely used medical imaging modality providing cross-sectional images of the patient used to detect pathological abnormalities. CT is used as a tool both for diagnosis and treatment planning for a wide range of disease such as stroke, cancer and cardiovascular disease. However, the potential risk associated with ionizing radiation \cite{degonzales2004, brenner2007} has spurred on a huge research endeavor to achieve images of high diagnostic quality while keeping the dose as low as reasonably achievable \cite{wang2020,koetzier2023}. Photon-counting CT (PCCT), based on the latest generation of CT detector technology, inherently contributes towards this objective as it is able to reduce dose via photon energy weighting and by largely eliminating the effects of electronic noise. This novel detector technology, in addition to improved low-dose imaging, yields major improvements in spatial and energy resolution \cite{willemink2018,flohr2020,danielsson2021,hsieh2021,higashigaito2022,rajendran2022} both extremely valuable to provide accurate diagnosis. However, obtaining high resolution in either space or energy decreases the number of photons in each respective voxel or energy bin, and this unavoidably increases image noise. Hence, to materialize the full potential of the latest in X-ray CT detector technology there is an even higher demand for high quality image denoising techniques. 

Existing image denoising techniques can roughly be categorized into: iterative reconstruction \cite{wang2006,thibault2007,sidky2008,yu2009,tian2011,stayman2013,zeng2017}, pre-processing methods \cite{lariviere2005}, and post-processing methods \cite{ma2011,zhang2016,chen2017,wolterink2017,yang2018, shan2019, kim2019, kim2020, yuan2020, makinen2020, li2021, wang2023, niu2023, liu2023,tivnan2023}. Iterative reconstruction have proved to be successful in generating images with low noise levels while keeping important details intact. However, these methods are usually computationally expensive. Pre-processing methods approach the problem in the sinogram domain, prior to image reconstruction. The advantage of this method is that it will be agnostic to specific parameters used in the image reconstruction (kernel, matrix size, field of view (FOV), etc.). However, as the sinogram is in general of higher dimension than the reconstructed image, these approaches impose a higher compute requirement and may simply be unfeasible in certain applications. Post-processing alleviates these issues by operating directly in the image domain. Popular post-processing methods include non-local means (NLM) \cite{ma2011,zhang2016} and block-matching 3D (BM3D) \cite{makinen2020} filtering, as well as deep learning-based methods\cite{chen2017, wolterink2017, yang2018, kim2019, shan2019, kim2020, yuan2020, li2021, wang2023, niu2023, liu2023, tivnan2023}. In particular, deep generative models have proved exceptionally capable in suppressing noise while preventing over-smoothing and thereby generating processed images with appealing noise characteristics \cite{wolterink2017,yang2018,liu2023,tivnan2023}. It is also possible to combine pre- and post-processing by considering both the image and sinogram domain within one method, as done in \cite{niu2022}.

Diffusion and Poisson flow models are relatively recent deep generative models that have shown excellent performance on a wide range of tasks, showing remarkable success for unconditional \cite{sohl-dickstein2015,ho2020,nichol2021,song2021,song2021b,karras2022,xu2022,xu2023} and conditional image generation \cite{song2021, batzolis2021,chung2021,song2021c,saharia2022b,saharia2023,chung2023,liu2023, tivnan2023}. These families of generative models, lend themselves very well for inverse problem solving, ubiquitous in medical imaging, and have already been demonstrated on a range of problems in medical imaging \cite{song2021c, liu2023, chung2023, tivnan2023, ge2023}. Despite being based on two widely different underlying physical processes, EDM\cite{karras2022} (diffusion models) and Poisson flow generative models (PFGM)++\cite{xu2023} are intimately connected in theory and in practice. The generative processes both work by iteratively denoising images, starting from an initial prior noise distribution, following some physically meaningful trajectory. The former is inspired by non-equilibrium thermodynamics and the latter by electrostatics. PFGM++ realize a generative model by treating $N$-dimensional data as electric charges in a $N+D$-dimensional augmented space. Tracing out the resulting electric field lines yields a trajectory, defined by an ordinary differential equation (ODE), from an easy-to-sample prior distribution to the data distribution of interest. Amazingly, the training and sampling processes of PFGM++ converges to that of EDM in the $D\rightarrow\infty, r=\sigma \sqrt{D}$ limit \cite{xu2023}. In other words, PFGM++ contains diffusion models as a special case. In addition, EDM and PFGM++ are also tightly connected in practice. As show in \cite{xu2023}, the training and sampling algorithms introduced for EDM \cite{karras2022} can directly be applied to PFGM++ with just an updated prior noise distribution and a simple change of variables. 

The iterative sampling process is a key feature of diffusion-style models, such as diffusion and Poisson flow models. This allows for a flexible trade-off between compute and image quality as well as zero-shot editing of data. However, this is also a key limitation as more compute means slower sampling which may limit their use in real-time applications. Compared to single-step models such as GANs \cite{goodfellow2014}, diffusion-style models may required on the order of $10-10^3$ times more compute to generate a sample, both for unconditional and conditional generation. Considering clinical CT image denoising as an example, a full 3D volume may contain hundreds of slices that promptly need to be processed. Efforts to reduce the number of function evaluations (NFE), and improve sampling speeds, include moving to efficient ODE samplers \cite{song2021b} and distillation techniques \cite{salimans2022}. A recent development is consistency models \cite{song2023}, which builds upon of probability flow diffusion models and learns to map any point at any time-step to the trajectory's initial point. This is achieved by enforcing self-consistency: any two points on the same trajectory maps to the same initial point. A consistency model can be trained in distillation mode (consistency distillation), where a pretrained diffusion model is distilled into a single-step sampler, and in isolation mode (consistency training), where a consistency model is trained from scratch as a stand-alone model. Although yielding impressive results, there is a noticeable drop in performance when comparing the output from the consistency model with NFE=1 to the underlying diffusion model with NFE$>$1. This drop in performance is smaller for consistency distillation than for consistency training and can be mitigated by taking a few more steps in the sampling process.

In this paper, we propose a novel post-processing denoising method that exploits the added robustness afforded by choosing $D$ in the PFGM++ framework to achieve high image quality without the penalty of computationally costly sampling. The main contributions are as follows: 1) We present \textbf{p}osterior sampling \textbf{P}oisson \textbf{f}low generative \textbf{m}odels (PPFM), a novel framework for image denoising in low-dose and photon-counting CT that produces excellent image quality whilst keeping NFE=1. Using PFGM++\cite{xu2023}, originally developed for unconditional generation (noise-to-image), as starting point, we update the training and sampling processes, utilizing paired data to learn a conditional generator (image-to-image). Intuitively, instead of estimating an empirical electric field as in PFGM++ \cite{xu2023}, we exploit the additional information afforded by paired data to estimate a ``conditional'' empirical electric field, which defines a trajectory from the prior noise distribution to the posterior distribution of interest. While not strictly necessary in order to get a sample from the desired posterior, we additionally hijack and regularize the sampling process. Using this formulation we can choose the hyperparameters such that NFE=1. 2) We shed light on the benefits of using the PFGM++ framework with variable $D$ compared to diffusion models with $D\rightarrow \infty$ fixed for the task of image denoising. The corresponding posterior sampling method based on diffusion models is contained as a special case ($D\rightarrow \infty$) in our proposed method and our results indicate that the PFGM++ framework, with $D$ as an additional hyperparameter, yields significant performance gains. 3) We show that our proposed method outperforms current state-of-the-art diffusion-style models with NFE=1, consistency models \cite{song2023}. In addition to the state-of-the-art from the AI literature, we also compare our proposed method to previous popular supervised (RED-CNN \cite{chen2017}, WGAN-VGG \cite{yang2018}), and non-deep learning-based (BM3D \cite{makinen2020}) image denoising techniques. Our results indicate superior performance on clinical low-dose CT images and clinical images from a prototype photon-counting CT scanner developed by GE HealthCare, Waukesha \cite{almqvist2023}.

Code used for this paper is available at: \url{https://github.com/dennishein/cpfgmpp_PCCT_denoising}.

\section{Methods}
\subsection{Problem formulation}
The objective in this paper is to generate high-quality reconstructions $\bm{\hat{y}} \in \mathbb{R}^N$ of $\bm{y} \in \mathbb{R}^N$ from noise degraded $\bm{c} = \mathcal{F}(\bm{y}) \in \mathbb{R}^N,$ where $\mathcal{F}: \mathbb{R}^N \rightarrow \mathbb{R}^N$ denotes the noise degradation operator, including factors such as quantum noise \cite{chen2017}, and $N:= n \times n$. In the case of low-dose CT, $\bm{y}$ corresponds to the normal-dose CT (NDCT) and $\bm{c}$ to the low-dose CT (LDCT) image. In the case of photon-counting CT, $\bm{c}$ is the thin unprocessed slice and $\bm{y}$ is its noise suppressed counterpart. The problem of generating high-quality reconstructions $\bm{\hat{y}}$ of $\bm{y}$ from measurements $\bm{c}$ is typically ill-posed. It helpful to treat this as a statistical inverse problem, and we will assume that the data follow some prior distribution $\bm{y}\sim p(\bm{y})$. Our high-quality reconstruction is then a sample from the posterior $\bm{\hat{y}} \sim p(\bm{y}|\bm{c})$. This strategy for solving inverse problem is called posterior sampling. In this paper, $\bm{y}$ will be treated as ``ground truth'' despite the fact that it may contain noise and artifacts.

\subsection{Diffusion models}
Diffusion models \cite{sohl-dickstein2015,ho2020,nichol2021,song2021b,song2021,karras2022}, originally inspired by non-equilibrium thermodynamics, work by first slowly transforming the data distribution to a noise distribution by iteratively adding Gaussian noise, and subsequently learning to run the process in reverse, slowly removing the noise. Building on the continuous-time probability flow ODE formulation in \cite{song2021}, reference \cite{karras2022} describes this process as
\begin{equation}
    d\bm{x} = - \dot{\sigma}(t)\sigma(t)\nabla_{\bm{x}}\log p_{\sigma(t)} (\bm{x}) dt, 
    \label{edm_ode}
\end{equation}
where $\sigma(t)$ is a predefined, time-dependent, noise scale and $\nabla_{\bm{x}}\log p_{\sigma(t)} (\bm{x})$ is the time-dependent score function of the perturbed data distribution. Moving the ODE forward and backward in time nudges the sample away from and towards the data distribution, respectively. Crucially, the ODE in Eq. \eqref{edm_ode} only depends on the data distribution via the time-dependent score function, an estimate of which can be obtained by minimizing the weighted denoising score matching \cite{vincent2011} objective 
\begin{align}
    \mathbb{E}_{\sigma \sim p(\sigma)} &\mathbb{E}_{\bm{y}\sim p(\bm{y
    })} \mathbb{E}_{\bm{x}\sim p_{\sigma}(\bm{x}|\bm{y})} \notag \\ 
    & \left[ \lambda(\sigma) ||f_{\theta}(\bm{x},\sigma)-\nabla_{\bm{x}} \log p_{\sigma}(\bm{x}|\bm{y})||_2^2\right], 
\end{align}
where $\lambda(\sigma)$ is a weighting function, $p(\sigma)$ the training distribution of noise scales, $p(\bm{y})$ the data distribution, and $p_{\sigma}(\bm{x}|\bm{y})=\mathcal{N}(\bm{y},\sigma^2\bm{I})$ the Gaussian perturbation kernel, which samples perturbed data $\bm{x}$ from ground truth data $\bm{y}.$ Once equipped with this estimate, we can generate an image by drawing an initial sample from the prior noise distribution and solving Eq. \eqref{edm_ode} using some numeric ODE solver. 

\subsection{PFGM++}
Instead of estimating a time-dependent score function, as for score-based diffusion models, the objective of interest in PFGM++ is the high dimensional electric field
\begin{equation}
    \bm{E} (\bm{\tilde{x}}) = \frac{1}{S_{N+D-1}(1)} \int \frac{\bm{\tilde{x}}-\bm{\tilde{y}}}{||\bm{\tilde{x}}-\bm{\tilde{y}}||^{N+D}} p(\bm{y}) d\bm{y},
    \label{pfgmpp_E}
\end{equation}
where $p(\bm{y})$ is the ground truth data distribution, $S_{N+D-1}(1)$ is the surface area of the unit $(N+D-1)$-sphere, and $\tilde{\bm{y}}:=(\bm{y},\bm{0}) \in \mathbb{R}^{N+D}$ and $\tilde{\bm{x}}:=(\bm{x},\bm{z}) \in \mathbb{R}^{N+D}$ the augmented ground truth and perturbed data, respectively. The electric field lines, generated by the data treated as electric charges in the augmented space, define a surjection between the ground truth data distribution and a uniform distribution on the the infinite $N+D$-dimensional hemisphere. Importantly, the electric field is rotationally symmetric on the $D$-dimensional cylinder $\sum_{i=1}^D z_i^2 = r^2, \forall r > 0$ and therefore a dimensionality reduction is possible \cite{xu2023}. In particular, it suffices to track the norm of the augmented variables $r = r(\bm{\tilde{x}}):=||\bm{z}||_2$ and we can redefine $\tilde{\bm{y}}:=(\bm{y},0) \in \mathbb{R}^{N+1}$ and $\tilde{\bm{x}}:=(\bm{x},r) \in \mathbb{R}^{N+1}$. Hence, the ODE of interest is 
\begin{equation}
    d\bm{x} = \bm{E} (\bm{\tilde{x}})_{\bm{x}} \cdot E(\bm{\tilde{x}})_r^{-1} \label{pfgmpp_ode} dr, 
\end{equation}
where $\bm{E}(\bm{\tilde{x}})_{\bm{x}}=\frac{1}{S_{N+D-1}(1)} \int \frac{\bm{x}-\bm{y}}{||\bm{\tilde{x}}-\bm{\tilde{y}}||^{N+D}}p(\bm{y})d\bm{y}$, and $E(\bm{\tilde{x}})_r=\frac{1}{S_{N+D-1}(1)} \int \frac{r}{||\bm{\tilde{x}}-\bm{\tilde{y}}||^{N+D}}p(\bm{y})d\bm{y}$, a scalar. Crucially, this symmetry reduction has converted the aforementioned surjection into a bijection between the ground truth data placed on the $r=0$ $(\bm{z}=\bm{0})$ hyperplane and a distribution on the $r=r_{\max}$ hyper-cylinder \cite{xu2023}. PFGM++ employs a perturbation based objective, akin to the denoising score matching objective in score-based diffusion models \cite{karras2022,song2021}. In particular, for the perturbation kernel $p_r(\bm{x} | \bm{y})$, the objective is
\begin{equation}
    \mathbb{E}_{r\sim p(r)} \mathbb{E}_{\bm{y}\sim p(\bm{y})} \mathbb{E}_{\bm{x} \sim p_r(\bm{x}|\bm{y})} \left[ ||f_\theta (\bm{\tilde{x}})-\frac{\bm{x}-\bm{y}}{r/\sqrt{D}} ||_2^2\right] 
    \label{pfgmpp_obj_final}
\end{equation}
where $p(r)$ the training distribution over $r$. The key idea is that we can choose the perturbation kernel such that the minimizer of Eq. \eqref{pfgmpp_obj_final} matches Eq. \eqref{pfgmpp_ode}. In particular, for $p_r(\bm{x}|\bm{y}) \propto 1/(|| \bm{x}-\bm{y}||_2^2+r^2)^{\frac{N+D}{2}}$, it is possible to show that the minimizer of Eq. \eqref{pfgmpp_obj_final} is $f^*_\theta(\bm{\tilde{x}}) = \sqrt{D} \bm{E} (\bm{\tilde{x}})_{\bm{x}} \cdot E(\bm{\tilde{x}})_r^{-1}.$ Starting with an initial sample from $p_{r_{\max}}$ one can generate a sample for the target data distribution by solving $d\bm{x}/dr = \bm{E}(\bm{\tilde{x}})_{\bm{x}}/E(\bm{\tilde{x}})_r = f^*_\theta(\bm{\tilde{x}})/\sqrt{D}$ using some numeric ODE solver.  

\subsection{Posterior sampling Poisson flow generative models}
Our proposed method, PPFM, builds on PFGM++, by updating both the training and sampling processes. There are many ways to obtain a conditional generator for diffusion models, as shown in \cite{batzolis2021}. The most straightforward of which is to simply feed the condition image $\bm{c}$ as an additional input to the network estimating the time-dependent score function. This has been used with great success empirically \cite{saharia2022b,saharia2023} and \cite{batzolis2021} showed mathematically that this ``trick'' has a solid theoretical background and does yield a consistent estimator of the conditional time-dependent score function. We will move from an unconditional generator to a conditional one following this strategy. For conciseness, we will leave a theoretical treatment to future work and instead illustrate empirically that this adjusted objective generates samples from the desired posterior. In practice, as is the case for PFGM++, we will employ the training and sampling algorithms from EDM \cite{karras2022 } using an updated prior noise distribution, the $r=\sigma \sqrt{D}$ hyperparameter translation formula, $\bm{\tilde{x}}:=(\bm{x},r)$, and the fact that EDM sets $\sigma(t)=t.$ Since $dr=d\sigma \sqrt{D}=dt \sqrt{D}$, by a change of variable we have that $d\bm{x} = f^*_\theta(\bm{\tilde{x}})/\sqrt{D} dr = f^*_\theta(\bm{\tilde{x}}) dt.$ The training process of PPFM is presented in Algorithm \ref{train_pfgmpp} with updates to the original formulation in PFGM++\cite{xu2023} highlighted in blue\footnote{Note that $f_\theta$ is estimated indirectly via $D_\theta.$}.

\begin{algorithm}
\DontPrintSemicolon
\SetAlgoNoLine
\textcolor{blue}{Sample data $\{\bm{y}_i, \bm{c}_i \}_{i=1}^{\mathcal{B}}$ from $p(\bm{y},\bm{c})$}\;
Sample standard deviations $\{\sigma_i \}_{i=1}^{\mathcal{B}}$ from $p(\sigma)$\;
Sample $r$ from $p_r$: $\{r_i=\sigma_i \sqrt{D} \}_{i=1}^{\mathcal{B}}$ \;
Sample radii $\{R_i=p_{r_i}(R) \}_{i=1}^{\mathcal{B}}$ \;
Sample uniform angles $\{\bm{v}_i=\frac{\bm{u}_i}{|| \bm{u}_i||_2} \}_{i=1}^{\mathcal{B}}, \bm{u}_i \sim \mathcal{N} (\bm{0},\bm{I})$ \;
Get perturbed data $\{\hat{\bm{y}}_i = \bm{y}_i + R_i \bm{v}_i\}_{i=1}^{\mathcal{B}}$\;
Calculate loss $\ell(\bm{\theta})= \sum_{i=1}^{\mathcal{B}} \lambda (\sigma_i) ||D_{\bm{\theta}}(\hat{\bm{y}}_i,\sigma_i,\textcolor{blue}{\bm{c}_i})-\bm{y}_i||_2^2$\;
Update network parameters $\bm{\theta}$ using Adam \; 
\caption{Proposed PPFM training. Adapted from PFGM++\cite{xu2023} with adjustments highlighted in blue.}
\label{train_pfgmpp}
\end{algorithm}

Formally, the updates in Algorithm \ref{train_pfgmpp} are sufficient to get a conditional generator. However, we found that additionally updating the sampling process can yield significant improvements in terms of sampling speed, a key issue for diffusion-style models. Hence, we propose to hijack and regularize the sampling process. Instead of running all the way from a sample from the prior noise distribution, we will hijack the sampling process at some $i=\tau \in \mathbb{Z}_+, \tau < T$ by simply inserting our condition image $\bm{x}_\tau =\bm{c}.$ Consequently, the for-loop will then run from $i=\tau$ instead of $i=0.$ With this additional hyperparameter $\tau$ we have that $\text{NFE}=2\cdot(T-\tau)-1,$ where $T$ is the total number of steps, or noise-scales. Initial results injecting a forward diffused condition image using the Gaussian perturbation kernel, as in done in e.g., \cite{chung2021} for diffusion models, did not seem to improve the results whilst introducing additional stochasticity. Thus we decided to go with this more simplistic, and novel, approach of directly injecting the condition image $\bm{c}.$ Since $T$ is inversely proportional to the step-size employed in the ODE solver, choosing a small $T$ is equivalent to setting a large step-size. This means that we get quite aggressive denoising but it comes at the cost of a larger local error as the local error using the $2^{\textrm{nd}}$ order method scales as $\mathcal{O}(h^3)$ with step size $h.$ As noted in \cite{xu2023}, PFGM++ is relatively less sensitive to step-size than EDM \cite{karras2022} and our results will show that using PFGM++ framework allows us to push the hyperparameters to an extreme where we have a large step-size yet achieve good performance. Finally, we add a regularization step. The particular regularizer used will depend on the inverse problem at hand. Since we are here interested in image denoising, simply applying the identity map suffices. Initial results using a low-pass filtered version of $\bm{x}_{\tau}$, as in e.g., \cite{chung2023} for diffusion models, did not improve performance. Hence, we opted to go with this more simplistic formulation. In other words, we will mix $\bm{x}_{i+1}$ with $\bm{x}_{\tau}=\bm{c}$, the input image we seek to denoise, using weight $w \in [0,1].$ Our proposed PPFM sampling is shown in Algorithm \ref{our_algo}, again with updates to PFGM++ \cite{xu2023} highlighted in blue. Together, Algorithm \ref{train_pfgmpp} and \ref{our_algo} yields our proposed method, PPFM. 

\begin{algorithm}
\DontPrintSemicolon
\SetAlgoNoLine
\textcolor{blue}{Get initial data $\bm{x}_\tau=\bm{c}$}\;
\For{$i= \textcolor{blue}{\tau},...,T-1$}{
    $\bm{d}_i = (\bm{x}_i-D_\theta(\bm{x}_i,t_i,\textcolor{blue}{\bm{c}}))/t_i$ \; 
    $\bm{x}_{i+1}=\bm{x}_i+(t_{i+1}-t_i)\bm{d}_i$ \; 
    \If{$t_{i+1}>0$}{
        $\bm{d}_i' = (\bm{x}_{i+1}-D_\theta (\bm{x}_{i+1},t_{i+1},\textcolor{blue}{\bm{c}}))/t_{i+1}$ \;
        $\bm{x}_{i+1}=\bm{x}_i+(t_{i+1}-t_i)(\frac{1}{2}\bm{d}_i+\frac{1}{2}\bm{d}_i')$ \;
    }
    $\textcolor{blue}{\bm{x}_{i+1} = w\bm{x}_{i+1}+(1-w)\bm{x}_{\tau}}$ \;
}
\Return $\bm{x}_T$
\caption{Proposed PPFM sampling. Adapted from PFGM++\cite{xu2023} with adjustments highlighted in blue.}
\label{our_algo}
\end{algorithm}

\section{Experiments}
\subsection{Datasets}
\subsubsection{Mayo low-dose CT data}
The dataset from the Mayo Clinic, used in the AAPM low-dose CT grand challenge \cite{aapm2017}, is used for training and validation. This publicly available clinical dataset contains images from 10 patients reconstructed using two different kernels and two different slice thicknesses on a $512\times512$ pixel grid. In this paper, we use the data with slice thickness 1~mm and reconstruction kernel D30 (medium). We split the data into a training set containing the first 8 patients, with a total of 4800 slices, and a validation set containing the final 2 patients with a total of 1136 slices. 

\subsubsection{Photon-counting CT data}
For test data we use images gathered as a part of a clinical study of a GE prototype photon-counting system \cite{almqvist2023}. The patients were scanned at Karolinska Insitutet, Stockholm, Sweden (Case 1, effective diameter 28 cm, $\mathrm{CDTI_{vol}}=10.12 \; \mathrm{mGy}$) and at the University of Wisconsin–Madison, Madison, WI (Case 2, effective diameter 36~cm, $\mathrm{CDTI_{vol}}=27.64~ \mathrm{mGy}$) with parameters listed in Table \ref{scan_table}. We reconstructed 70~keV virtual monoenergetic images with filtered backprojection on a $512 \times 512$ pixel grid with 0.42~mm slice thickness.

\begin{table}
    \centering
    \begin{tabular}{ccccc} \toprule
	 & Tube current & Helical pitch & Rotation time & kVp \\ \midrule 
    Case 1 & 255~mA & 0.990:1 & 0.6~s & 120 \\ 
    Case 2 & 290~mA & 0.510:1 & 0.7~s & 120 \\ 
 \bottomrule \\
    \end{tabular}
    \caption{Key parameters used for scanning patients on prototype photon-counting CT system. The PCCT data are used for testing only.}
    \label{scan_table} 
\end{table}

\subsection{Implementation details}
We train a network for each $D \in \{64,128\}$ and for $D\rightarrow \infty$ for 100k iterations using Adam \cite{kingma2014} with learning rate $2\times10^{-4}$ and batch size of 32 on one NVIDIA A6000 48GB GPU. We borrow the majority of the hyperparameters directly from \cite{xu2023}. We use DDPM++ with channel multiplier 128, channels per resolution [1, 1, 2, 2, 2, 2, 2], and self-attention layers at resolutions 16, 8, and 4. The only adjustment to the network architecture to move from a unconditional to a conditional generator, is to adjust the number of channels. The suggested preconditioning, exponential moving average (EMA) schedule, and non-leaky augmentation from \cite{karras2022} is used with an augmentation probability of $15\%$. We in addition set dropout probability to $10\%.$ The network is trained on randomly extracted $256 \times 256$ patches. Training on patches will lead to efficient training (lower graphics memory requirements) and additionally help prevent overfitting as training on randomly extracted patches serves as additional data augmentation. We train the network using mixed precision to further reduce the graphics memory requirements. $\tau$,~$T$ and $w$ are crucial hyperparameters in Algorithm \ref{our_algo}. As we only consider setups with NFE=1 for our main results, $\tau=T-1$ and hence completely determined by $T.$ We set $T$ and $w$ by grid search over $T\in \{4,8,16,32,64\}$ and $w\in \{0.5,0.6,0.7,0.8,0.9,1.0\}$ using Learned Perceptual Image Patch Similarity (LPIPS) \cite{zhang2018} on the validation set as selection criteria for each $D \in \{64,128\}$ and for $D\rightarrow \infty$. This yields $T=8$ and $w=0.7.$ We note that even though NFE=1, this ``single-step'' configuration will also blend in the condition image, a second step. However, the time required for this operation is negligible and thus we still refer to this a ``single-step.'' 

\subsection{Comparison to other methods}
Consistency models \cite{song2023} are the current state-of-the-art diffusion-style models with NFE=1. However, as for the case of EDM \cite{karras2022} and PFGM++\cite{xu2023}, the original formulation is for the problem of unconditional image generation. To the best of our knowledge, consistency models have never been used for conditional generation. Nevertheless, since they build upon diffusion models, and the consistency distillation approach in particular distills said diffusion model to a consistency model, one can reasonably surmise that the strategy of feeding the condition images $\bm{c}$ as additional input to the network to get a conditional generator will work well. Our empirical results support this hypothesis. Starting from the official implementation\footnote{\url{https://github.com/openai/consistency_models}.} we employ minimal adjustments in order to learn a conditional consistency model with $\bm{c}$ as additional input using the, consistency distillation approach. We opt for the consistency distillation, instead of consistency training, as this is the top performing approach in \cite{song2023}. We train the networks on randomly extracted $256\times256$ patches from the 8 patients in the Mayo low-dose CT training data. All hyperparameters for training and sampling are set as in \cite{song2023} for the LSUN $256\times 256$ experiments\footnote{As specified in \url{https://github.com/openai/consistency_models/blob/main/scripts/launch.sh}.}, except for batch size with had to be reduced to 4 to fit on a single NVIDIA A6000 48GB GPU. We first train an EDM for 300k iterations, and subsequently distill it into a consistency model during 600k iterations. For data augmentation, we applied random rotations and mirrorings. It is worth pointing out that this network has approximately a factor 11 more learnable parameters than what we use for our proposed method. In addition, it is trained for considerably more iterations. Hence, both sampling and training are considerably more time consuming. In particular, despite both achieving NFE=1, our proposed PPFM offer 3.5 times faster sampling. Following \cite{song2023}, we will refer to this consistency model as CD (consistency distillation). 

In addition to the state-of-the-art from the AI literature, we also compare our proposed method to previous popular supervised and non-deep learning-based image denoising techniques. As an example of a popular non-deep learning-based technique we use a version of BM3D\cite{makinen2020}. BM3D was shown to be the top performer for Mayo low-dose CT denoising in the category of non-deep learning-based image denoising techniques in \cite{chen2017}. We used bm3d.py\footnote{\url{https://pypi.org/project/bm3d/}.} and set the parameter $\sigma_{\text{BM3D}}$ equal to the standard deviation of a flat region-of-interest (ROI) in the low-dose CT validation data. For supervised techniques we use RED-CNN \cite{chen2017} and WGAN-VGG \cite{yang2018}. RED-CNN was trained on over $10^6$ extracted overlapping $55 \times 55$ patches from the 8 patients in the Mayo low-dose CT training data. The architecture is set as specified in \cite{chen2017}. WGAN-VGG was trained on randomly extracted $64\times64$ patches from the training set, with network architecture and other hyperparameters as in \cite{yang2018}. For both networks, we augment the data by applying random rotations and mirrorings during training. WGAN-VGG is an interesting comparison case as it is very similar in principle to the method proposed in this paper. Both methods achieve image denoising via posterior sampling by adjusting the training processes of deep generative models, and thereby acquire conditional generators. The major difference is the deep generative model itself. WGAN-VGG \cite{yang2018} is based on GANs, which were the state-of-the-art deep generative models until the event of diffusion models and PFGM++, whereas our proposed method is based on PFGM++, a current state-of-the-art deep generative model. Despite being similar in principle, this difference leads to a myriad of important differences in practice. Notably, PFGM++ does not require adversarial training and is therefore much more stable to train. 

\subsection{Evaluation methods}
In addition to image quality assessment via visual inspection, we also consider three quantitative metrics of image quality. We employ the two most commonly used metrics in the CT denoising literature, namely structural similarity index (SSIM \cite{wang2004}) and peak signal-to-noise ratio (PSNR). These metrics are easy to use and very well established but they do not necessarily correlate well with human perception \cite{zhang2018}. PSNR is inversely proportional to the $\ell_2$ Euclidean distance. This simple pixel-wise metrics does not adequately capture nuances of human perception. This is particularly most evident for the case of blurring as a result of over-smoothing, which is inadequately penalized. On the other hand, SSIM is perceptually motivated; however, it is very difficult to model the complex processes underlying human perception and therefore is also falls short. Reference \cite{zhang2018} suggest using pretrained convolutional neural networks (CNNs) as feature extractors, as is the case for perceptual loss functions, to develop a metric of image similarity that closely corresponds to human perception. They call this metric LPIPS and demonstrate on a series of different datasets, using different pretrained CNNs, how LPIPS better corresponds to human perception than traditional metrics such as SSIM and PSNR. In this paper, we use the official implementation of LPIPS\footnote{\url{https://github.com/richzhang/PerceptualSimilarity}.} with AlexNet \cite{krizhevsky2014} as feature extractor.

\subsection{Results}
\begin{figure}
    \centering
    \includegraphics[width=\columnwidth]{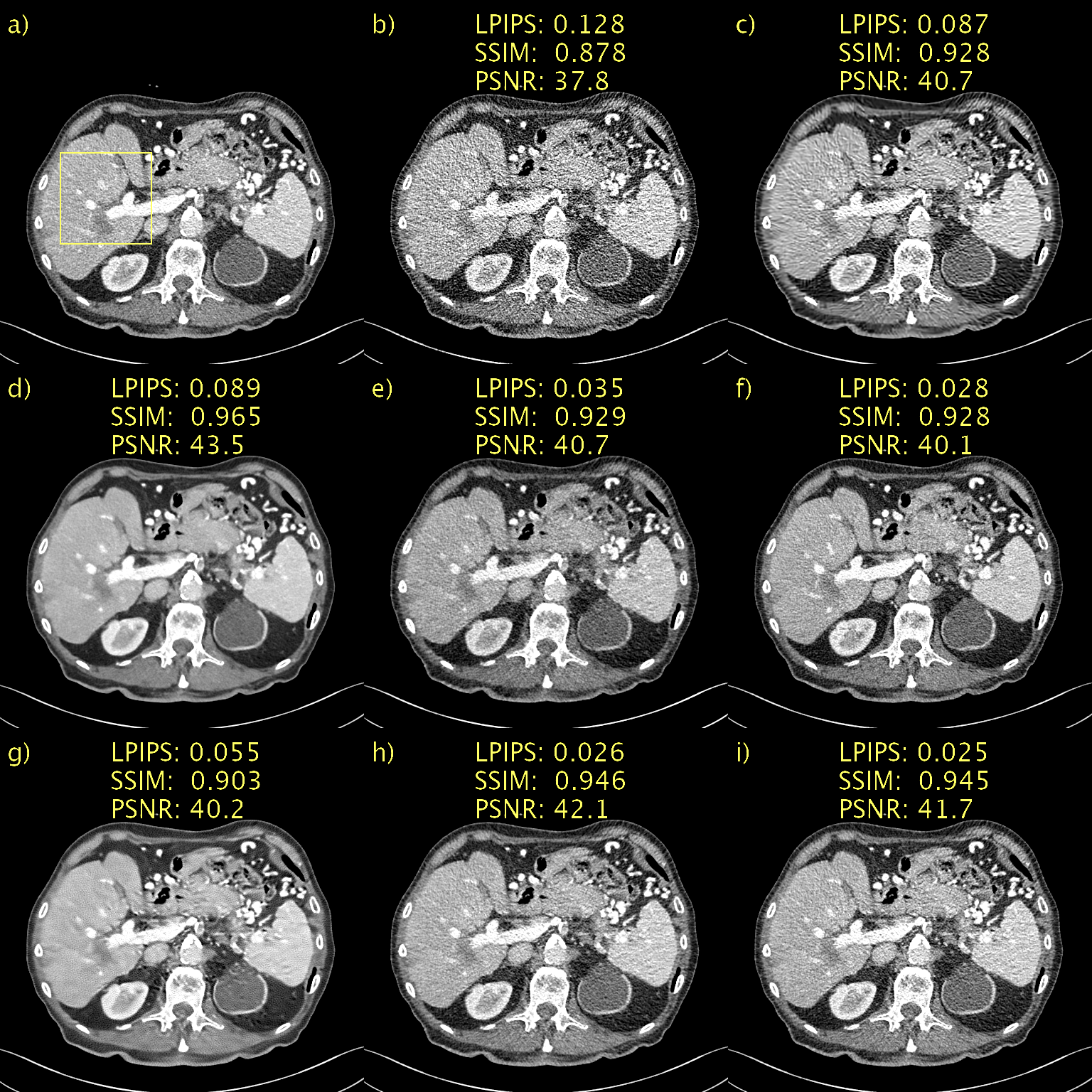}
    \caption{Results on the Mayo low-dose CT validation data. Abdomen image with a metastasis in the liver. a) NDCT, b) LDCT, c) BM3D \cite{makinen2020}, d) RED-CNN\cite{chen2017}, e) WGAN-VGG\cite{yang2018}, f), CD \cite{song2023}, g) PPFM ($D\rightarrow \infty$), h) PPFM ($D=128$), i) PPFM ($D=64$). Yellow box indicating ROI shown in Fig. \ref{1049_ex}. 1~mm-slices. Window setting [-160,240]~HU.}
    \label{1049}
\end{figure}

\begin{figure}
    \centering
    \includegraphics[width=\columnwidth]{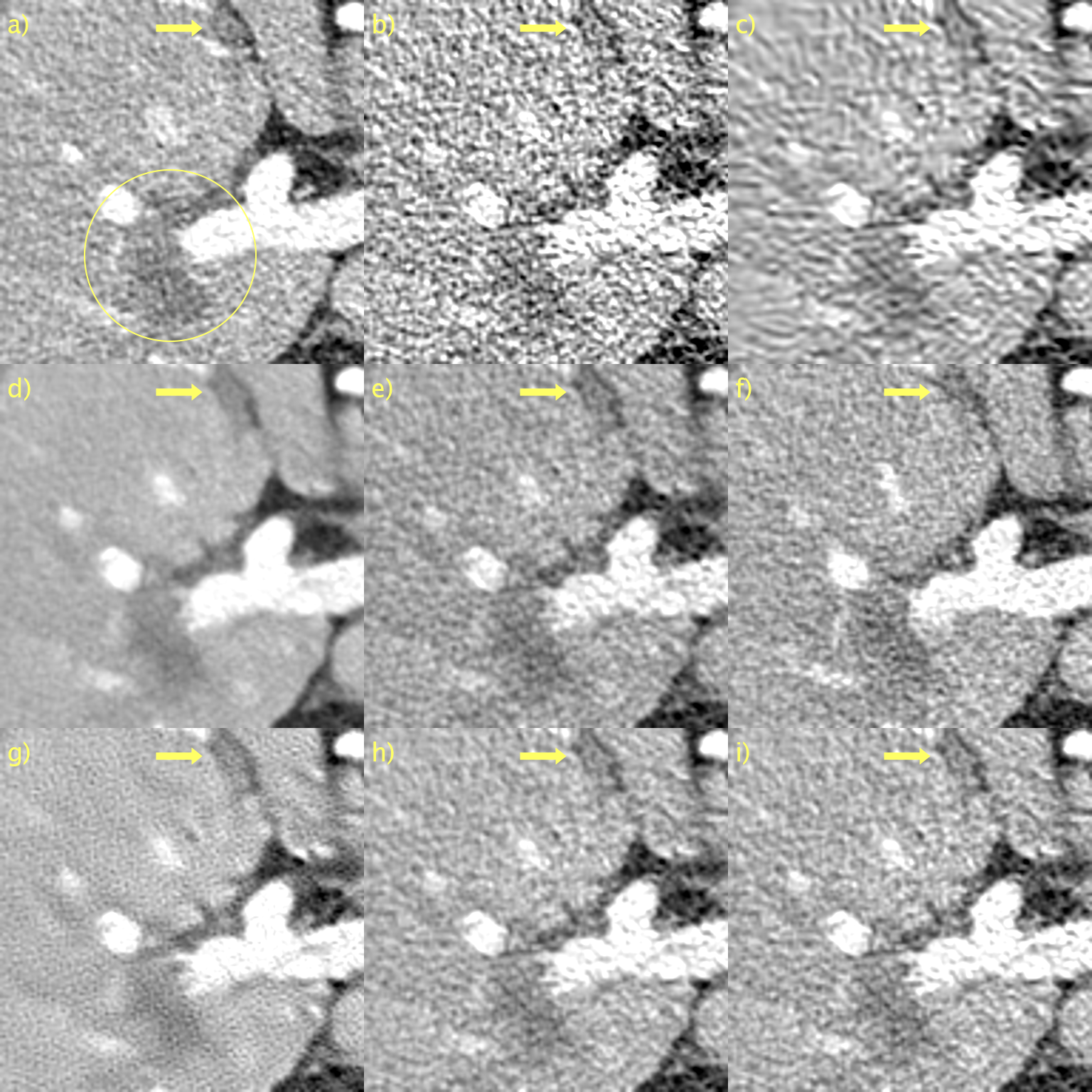}
    \caption{ROI in Fig. \ref{1049} magnified to emphasize details. a) NDCT, b) LDCT, c) BM3D \cite{makinen2020}, d) RED-CNN\cite{chen2017}, e) WGAN-VGG\cite{yang2018}, f), CD \cite{song2023}, g) PPFM ($D\rightarrow \infty$), h) PPFM ($D=128$), i) PPFM ($D=64$). Yellow circle added to emphasize lesion. Yellow arrow placed to emphasize detail. 1~mm-slices. Window setting [-160,240]~HU.}
    \label{1049_ex}
\end{figure}

\begin{figure*}
    \centering
    \includegraphics[width=\linewidth]{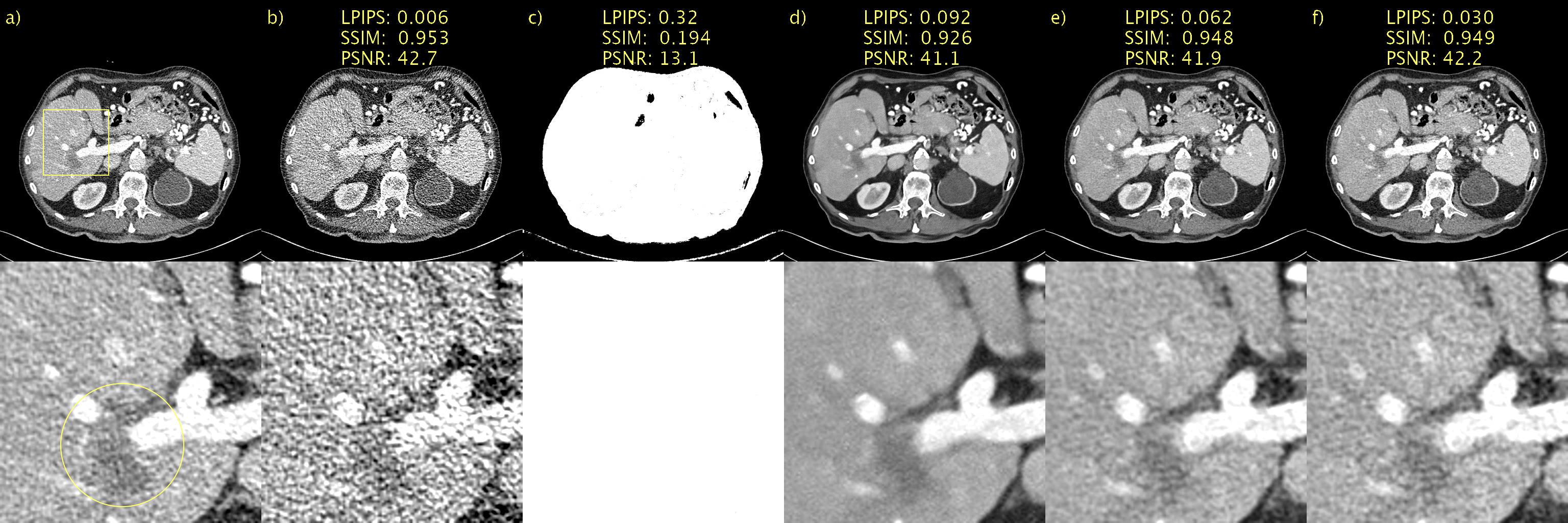}
    \caption{Results without hijacking and regularization. a) NDCT, b) LDCT, c) $T=8$, d) $T=16$, e) $T=32$, f) $T=64$. Yellow circle added to emphasize lesion. 1~mm-slices. Window setting [-160,240]~HU.}
    \label{18_ablation}
\end{figure*}

\begin{figure*}
    \centering
    \includegraphics[width=\linewidth]{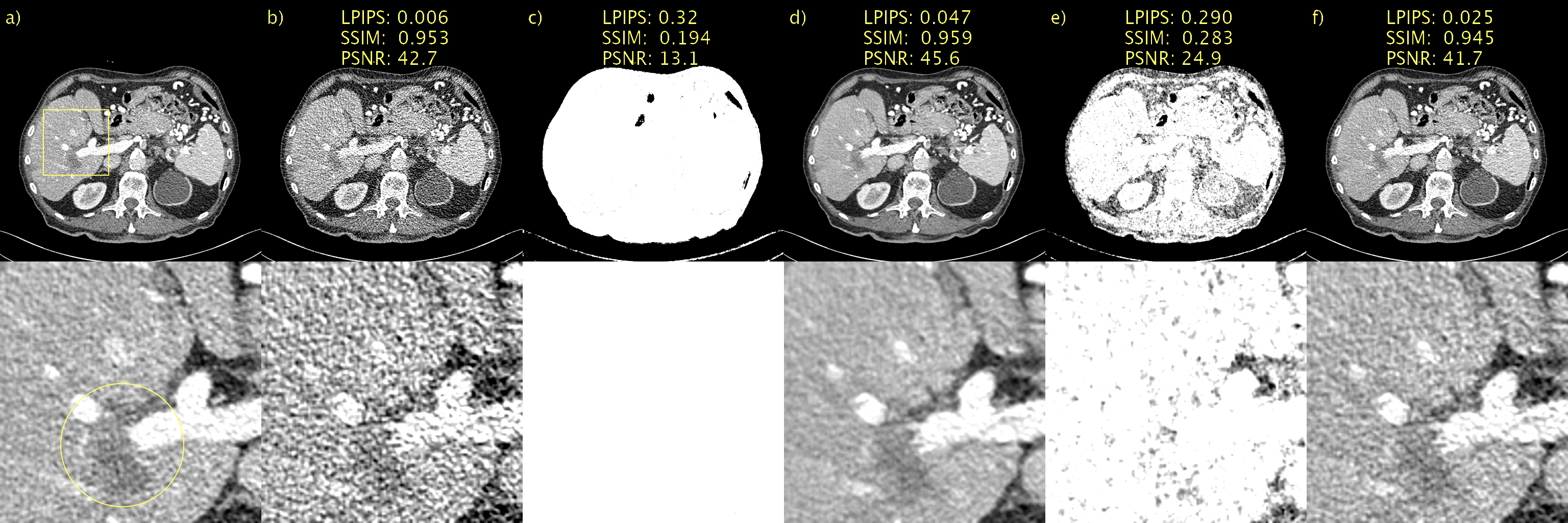}
    \caption{Ablation study of PPFM sampler. a) NDCT, b) LDCT, c) no hijacking and no regularization (NFE=7), d) hijacking but no regularization (NFE=1), e) no hijacking but regularization (NFE=7), f) hijacking and regularization (NFE=1). Yellow circle added to emphasize lesion. 1~mm-slices. Window setting [-160,240]~HU.}
    \label{18_ablation_alt}
\end{figure*}

\begin{figure}
    \centering
    \includegraphics[width=\columnwidth]{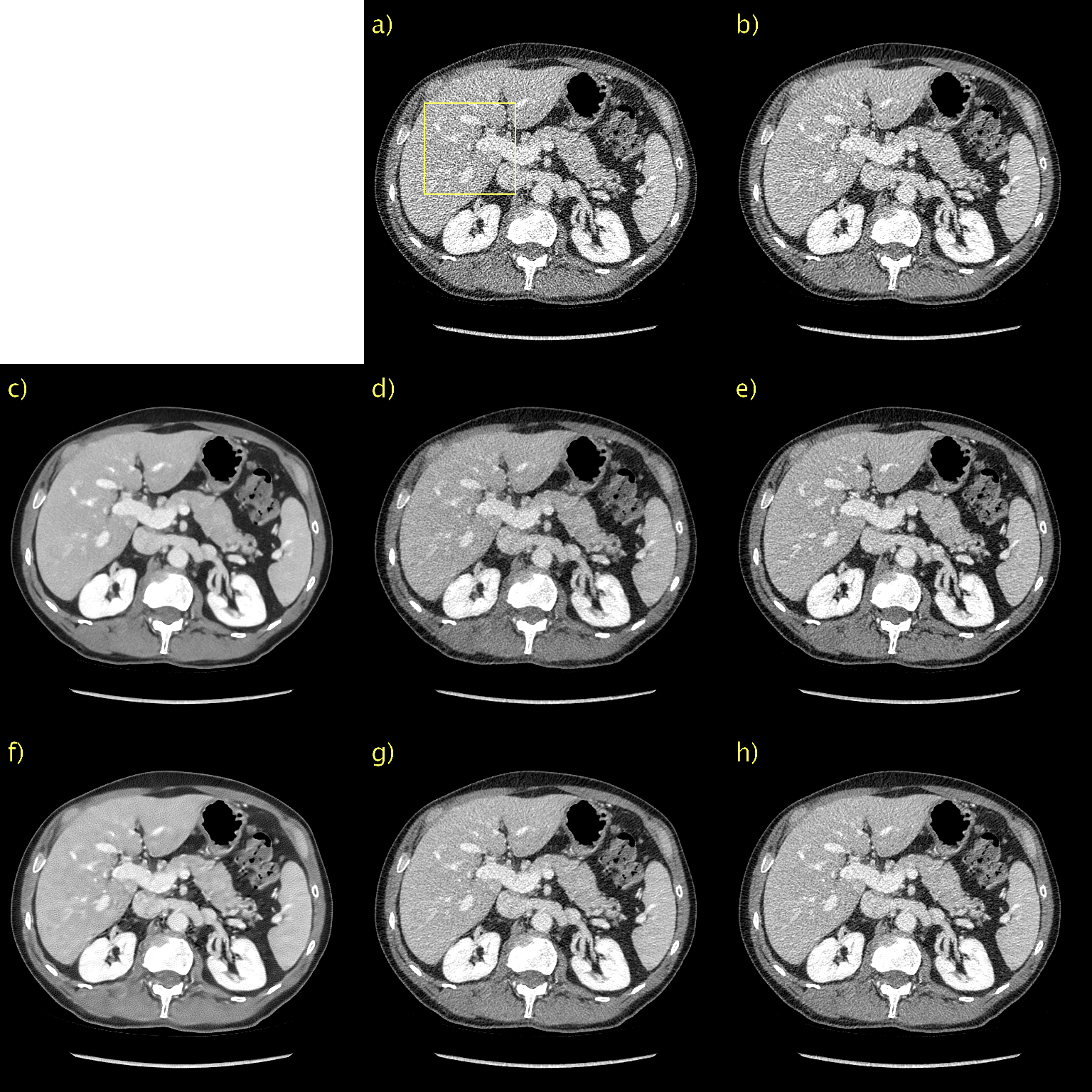}
    \caption{Results for the PCCT test data: Case 1. a) Unprocessed, b) BM3D \cite{makinen2020}, c) RED-CNN\cite{chen2017}, d) WGAN-VGG\cite{yang2018}, e), CD \cite{song2023}, f) PPFM ($D\rightarrow \infty$), g) PPFM ($D=128$), h) PPFM ($D=64$). No ground truth available. Yellow box indicating ROI shown in Fig. \ref{233_ex}. 0.42~mm-slices. Window setting [-160,240]~HU.}
    \label{233}
\end{figure}

\begin{figure}
    \centering
    \includegraphics[width=\columnwidth]{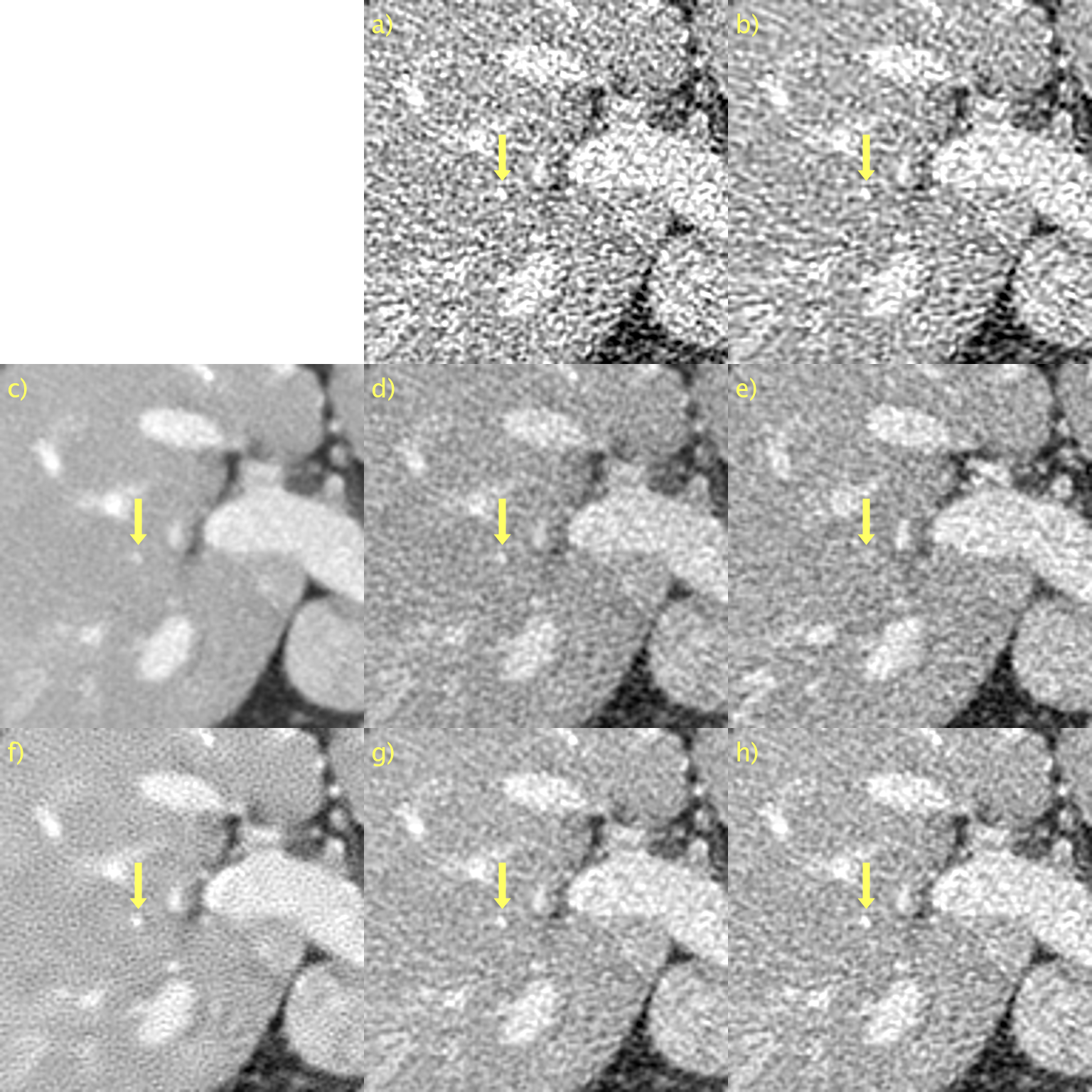}
    \caption{ROI in Fig. \ref{233} magnified to emphasize details. a) Unprocessed, b) BM3D \cite{makinen2020}, c) RED-CNN\cite{chen2017}, d) WGAN-VGG\cite{yang2018}, e), CD \cite{song2023}, f) PPFM ($D\rightarrow \infty$), g) PPFM ($D=128$), h) PPFM ($D=64$). No ground truth available. Yellow arrow placed to emphasize detail. 0.42~mm-slices. Window setting [-160,240]~HU.}
    \label{233_ex}
\end{figure}

\begin{figure}
    \centering
    \includegraphics[width=\columnwidth]{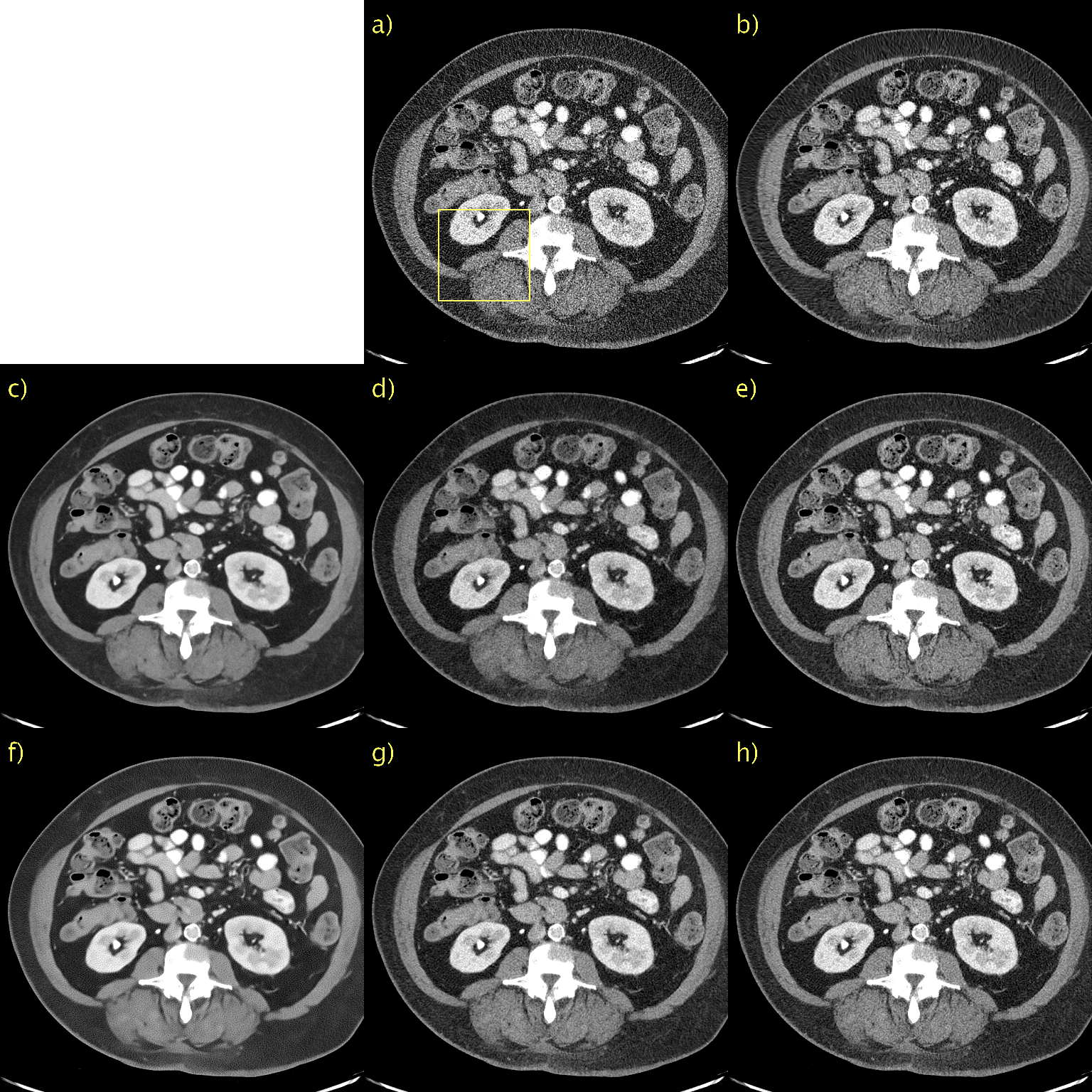}
    \caption{Results for the PCCT test data: Case 2. a) Unprocessed, b) BM3D \cite{makinen2020}, c) RED-CNN\cite{chen2017}, d) WGAN-VGG\cite{yang2018}, e), CD \cite{song2023}, f) PPFM ($D\rightarrow \infty$), g) PPFM ($D=128$), h) PPFM ($D=64$). No ground truth available. Yellow box indicating ROI shown in Fig. \ref{317_ex}. 0.42~mm-slices. Window setting [-160,240]~HU.}
    \label{317}
\end{figure}

\begin{figure}
    \centering
    \includegraphics[width=\columnwidth]{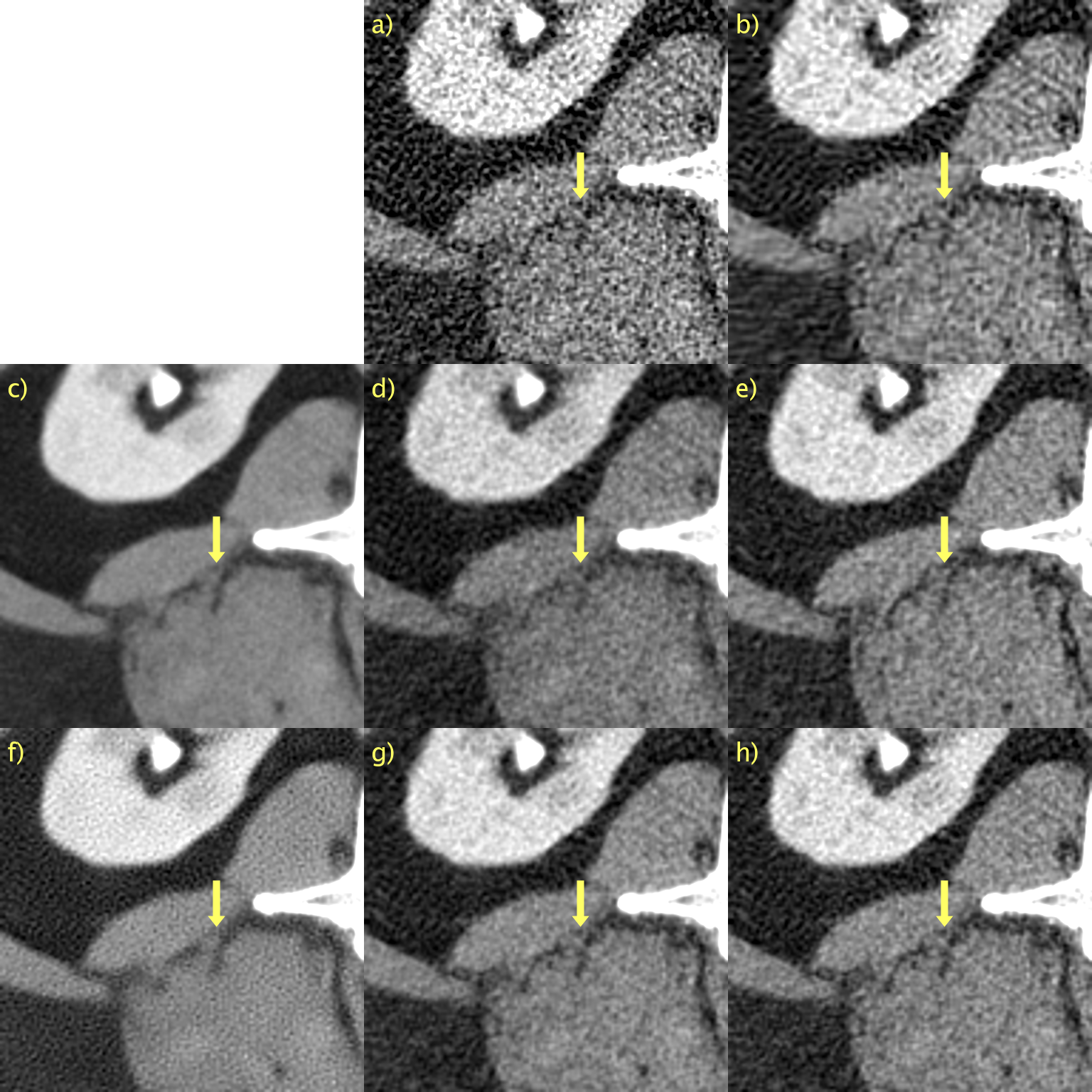}
    \caption{ROI in Fig. \ref{317} magnified to emphasize details. a) Unprocessed, b) BM3D \cite{makinen2020}, c) RED-CNN\cite{chen2017}, d) WGAN-VGG\cite{yang2018}, e), CD \cite{song2023}, f) PPFM ($D\rightarrow \infty$), g) PPFM ($D=128$), h) PPFM ($D=64$). No ground truth available. Yellow arrow placed to emphasize detail. 0.42~mm-slices. Window setting [-160,240]~HU.}
    \label{317_ex}
\end{figure}

Qualitative results, along side with LPIPS, SSIM and PSNR, for a representative case from the Mayo low-dose CT validation data are available in Fig. \ref{1049} and \ref{1049_ex}. This patient is of additional interest due to a metastasis in the liver. To emphasize this lesion we include a magnified version of the ROI in Fig. \ref{1049} in Fig. \ref{1049_ex}. BM3D, shown in c), does a good job suppressing noise and recovering details. However, this comes at a cost of artifacts that makes the image appear smudgy. RED-CNN, shown in d), does an exceptional job of suppressing noise whilst keeping key details intact. Nevertheless, the denoising is too aggressive and the noise is suppressed well below the level in the NDCT image, shown in a). This over-smoothing is expected since RED-CNN is trained with a simple pixel-wise $\ell_2$-loss. WGAN-VGG, shown in e), on the other hand, does a very good job at suppressing noise while producing noise characteristics aligned with that of the NDCT image. At first glance, CD, shown in f), seems to perform exceedingly well. However, at closer inspection, especially in Fig. \ref{1049_ex}, one can see several details that appear different for CD than for all the other images, including NDCT and LDCT. We highlighted one such detail with a yellow arrow. CD seems to have added a feature that is not visible in the LDCT nor NDCT image. Seemingly convincing, but factually inaccurate, claims are commonly referred to as ``hallucinations'' in the large language models (LLMs) literature.\footnote{See, for instance, reference \cite{tian2023} for an overview.} We will adopt this terminology to mean inaccurate addition, or removal, of features. Results for our proposed method are available in g)-i). PPFM, with $D=128$ and $D=64,$ does an exceptional job of suppressing noise whilst keeping key details intact and accurately reproducing the noise characteristics of the NDCT image. Comparing g), with $D\rightarrow \infty$, to $D$ finite, in h) and i), emphasizes the effect of the added robustness afforded by choosing $D$ in PFGM++ framework. For small $T,$ or equivalently a large step-size, PPFM with $D\rightarrow \infty$ breaks down whereas PPFM with $D$ finite yield good results.\\ 
\indent The mean and standard deviation of LPIPS, SSIM, and PSNR over the entire Mayo low-dose CT validation set are available in Table \ref{table_quant}. The top performer in terms of SSIM and PSNR is RED-CNN. This is not entirely unexpected since RED-CNN is trained to minimize the $\ell_2$-loss between patches from the NDCT and LDCT images. However, as noted above, SSIM and PSNR do not necessarily correspond well with human perception---in particular when it comes to over-smoothing. WGAN-VGG combines a perceptual loss with an adversarial loss in order to generate a denoised image from a posterior that is ``close'', in a certain sense, to the distribution of the NDCT images. The overall noise characteristics, texture and level, more closely resembles that of the NDCT image for WGAN-VGG than for RED-CNN. We can see that, accordingly, the LPIPS is significantly lower (better) for WGAN-VGG than RED-CNN. The overall top performer in terms of LPIPS is our proposed method, PPFM, with $D=64.$ \\
\begin{table}
    \centering
    \begin{tabular}{cccc} \toprule
	& LPIPS ($\downarrow$)& SSIM ($\uparrow$)&	PSNR ($\uparrow$)\\ \midrule 
        LDCT   & $0.075 \pm 0.02$ & $0.94 \pm 0.02 $ & $41.5\pm 1.6$ \\
        BM3D\cite{makinen2020} & $0.050 \pm 0.01$ & $0.97 \pm 0.01 $ & $45.0\pm 1.6$\\
        RED-CNN\cite{chen2017} & $0.048 \pm 0.02$ & $\bm{0.98} \pm 0.01 $ & $\bm{46.8}\pm 1.2$\\
        WGAN-VGG\cite{yang2018} & $0.019 \pm 0.01$ & $0.96 \pm 0.01 $ & $43.2\pm 0.9$\\
        CD\cite{song2023} & $0.013 \pm 0.00$ & $0.96 \pm 0.01 $ & $43.1\pm 1.0$\\
        \midrule
        PPFM & & & \\
        \midrule 
        $D\rightarrow \infty$ & $0.025 \pm 0.01$ & $0.93 \pm 0.01 $ & $42.0\pm 0.7$\\
        $D=128$ & $0.012 \pm 0.00$ & $\bm{0.98} \pm 0.01 $ & $45.8\pm 1.4$\\
        $D=64$ & $\bm{0.010} \pm 0.00$ & $0.97 \pm 0.01 $ & $45.4\pm 1.4$\\ 
    \bottomrule \\
    \end{tabular}
    \caption{Mean and standard deviation of LPIPS, SSIM, and PSNR in the low-dose CT validation set. $\downarrow$ means lower is better. $\uparrow$ means higher is better. Best results in bold.}
    \label{table_quant} 
\end{table}
\indent The proposed method is trained in a supervised manner to directly yield a conditional estimator. Hence, as mentioned above, neither hijacking nor regularization is strictly necessary. Instead, one can simply draw an initial sample from the prior noise distribution and then solve the ODE to generate a sample from the desired posterior. To illustrate empirically that this is indeed the case, we set $\tau=0$, $w=1,$ and replace the first line with an initial sample from the prior noise distribution, $p_{r_{\max}},$ in Algorithm \ref{our_algo}. Hence, except for the fact that the network takes the condition image as an additional input, Algorithm \ref{our_algo} is exactly as in PFGM++ \cite{xu2023}. We show results for $T \in \{8,16,32,64\}$ in Fig. \ref{18_ablation}. Consistent with expectations, the performance improves as $T$, the total number of steps, gets larger. Crucially, we can see that for $T>32$ our high-quality reconstruction is a good approximation of the ground truth image, that is $\bm{\hat{y}}\approx \bm{y}.$ \\
\indent To shed light on the individual components of our proposed sampler, we conduct an ablation study with results available Fig. \ref{18_ablation_alt}. A) and b) show the NDCT and LDCT images, respectively. In c), we turn off hijacking and regularization. As was also seen in Fig. \ref{18_ablation}, the sampler breaks down in this setting. The same holds true in e), where we regularize but have turned off the hijacking. Comparing c) to d), we can see that hijacking plays a pivotal role in our proposed sampler. For the setting consider here, with $T=8$, hijacking allows us to move from a total breakdown to a very pleasing image. Regularizing is also shown to be beneficial, it helps prevent over-smoothing resulting from aggressive denoising, a consequence of choosing a large step-size, as can be seen when comparing d) to f). Hence, hijacking and regularization, hijacking in particular, is what enables excellent image quality whilst keeping NFE=1. In order words, hijacking can help break the dependence on large $T$ for good image quality. \\
\indent Results for a representative case from the PCCT test data, Case 1, are available in Fig. \ref{233} and \ref{233_ex}. Since these data are clinical images from a prototype photon-counting system, there are no images available to play the role of ``ground truth,'' and we will therefore have to resort to visual inspection as means of accessing image quality. Larger details can reasonably be distinguished from statistical variation in the noise; however, this is very difficult for smaller, lower contrast, details. With that caveat, since no ``ground truth'' is available, we simply define a good result as an image which preserves details visible in the unprocessed image, shown in a), but with a lower noise level. BM3D, shown in b), seems to generalize quite poorly. The noise level in b) similar to that in a) with additional artifacts that makes the image appear smudgy. This may be due to differences in noise characteristics in the validation data, where we measured $\sigma_{\text{BM3D}}$, and the test data. RED-CNN, WGAN-VGG, and CD, on the other hand, shown in c), d), and e), respectively, seem to generalize well from the low-dose CT data to the photon-counting CT test data. We have placed a yellow arrow on a detail of interest. This feature is clearly visible is all cases, including the unprocessed image, but it is missing for CD, shown in e). Hence, although it is difficult to say definitively without a ``ground truth,'' this seems to indicate that CD removed a genuine feature. The proposed method is shown in f)-h). As was the case for the Mayo low-dose CT validation data, there is a major performance boost for $D$ finite, shown in g) and h), compared to $D\rightarrow \infty$, shown in f). \\
\indent We show the results on the second PCCT test case in Fig. \ref{317}, with a magnified version of the ROI shown in Fig. \ref{317_ex}. We have also placed a yellow arrow in Fig. \ref{317_ex} to draw attention to specific details. We note that BM3D, shown in c), seems to be doing a better job in terms of noise suppression that in Fig. \ref{233}. Differences in performance in the two test cases is most likely due to differences in noise characteristics. RED-CNN, WGAN-VGG, and CD, shown in c), d), and e), respectively, all do a good job suppressing the noise while preserving details. The main difference is the characteristics, texture and level, of the resulting noise. In particular, RED-CNN, is notably very smooth. The $D\rightarrow\infty$ case, shown in f), over-smooths the image, reducing the contrast of key details, while introducing a strange texture. On the other hand, for finite $D$, PPFM results in images with realistic noise level and texture, and preserve key details. Moreover, we can see that the contrast of the fat in the back muscle, marked by the yellow arrow, is significantly better preserved using the proposed method with finite $D$, shown in g) and h), than for WGAN-VGG, shown in d). \\

\section{Discussion and conclusion}
It is likely the case that one achieve better performance using a multi-step sampler, trading off compute for image quality. Since we were here interested in the single-step case, only limited time was spend exploring the hyperparameters space for $\tau \neq T-1.$ In this preliminary search, we were unable to find a combination of $T, \tau$ and $w$ outperforming our current hyperparameters in terms of LPIPS on the Mayo low-dose CT validation set. It is left to future research to explore the extent to which there is a penalty in performance due to enforcing $\tau=T-1$, and thereby achieving NFE=1.\\
\indent Since we are interested in PCCT, the ultimate objective is to get an image denoising technique that works for spectral CT. Extending PPFM to the spectral case can be done in many different ways. One possibility is to simply expand the number of channels for each data point. Instead of feeding a single-energy image, one can use pairs of basis images or virtual monoenergetic images at two different energy levels. Assessing whether such an update would be sufficient, or if further updates are required to obtain a spectral CT denoiser is an interesting avenue for future research. \\
\indent Finally, this is a 2D image denoising method. As such, due to the nature of CT data, we are leaving an abundance of useful information on the table by not considering adjacent slices. We surmise that it should be relatively straight forward to extend the proposed method to a 3D denoiser and thus leave this to future work. \\ 
\indent In conclusion, we have presented PPFM, a novel image denoising technique for low-dose and photon-counting CT. Our proposed method updates the training and sample processes of PFGM++\cite{xu2023} to get an conditional generator which is able to achieve high image quality without the penalty of computationally costly sampling. In particular, our proposed method is a single-step sampler, that is NFE=1. Our results shed light on the benefits of building upon the PFGM++ framework, where $D$ is a tunable hyperparameter, compared to diffusion models where $D\rightarrow \infty$ is fixed. In particular, we demonstrate that the corresponding setup with a diffusion model fails. Our results demonstrate favorable performance compared to current state-of-the-art diffusion-style models with NFE=1, consistency models, as well as several popular deep learning-based and conventional post-processing techniques on clinical low-dose CT images and clinical images from a prototype photon-counting CT system.


\section*{Acknowledgment}
The authors declare the following financial interests/personal relationships which may be considered as potential competing interests with the work reported in this paper: DH discloses research collaboration with GE HealthCare. MP discloses research collaboration and license fees, GE HealthCare.

\end{document}